# Reconfigurable, non-volatile control of optical anisotropy in ReS$_2$ via ferroelectric gating

Mahfujur Rahaman[1,†], Seunguk Song[1,2,3,†,*], Aaliyah C. Khan[4], Bongjun Choi[1], Aaron M. Schankler[4], Kwan-Ho Kim[1], Wonchan Lee[2,3], Jason Lynch[1], Hyeon Suk Shin[3,5], Andrew M. Rappe[4], and Deep Jariwala[1]*

[1]*Department of Electrical and Systems Engineering, University of Pennsylvania, Philadelphia, Pennsylvania 19104, United States*
[2]*Department of Energy Science, Sungkyunkwan University (SKKU), Suwon 16419, Republic of Korea*
[3]*Center for 2D Quantum Heterostructures (2DQH), Institute for Basic Science (IBS), Suwon 16419, Republic of Korea*
[4]*Department of Chemistry, University of Pennsylvania, Philadelphia, Pennsylvania 19104, USA.*
[5]*Department of Chemistry, Sungkyunkwan University (SKKU), Suwon, 16419, Republic of Korea*

[†]These authors contributed equally: Mahfujur Rahaman, Seunguk Song
[*]Correspondence should be addressed. Email to: seunguk@skku.edu (S.S.), dmj@seas.upenn.edu (D.J.)

## ABSTRACT

Electrically tunable linear dichroism (LD) with non-volatile properties represents a critical yet elusive feature for next-generation integrated photonic elements in practical device architectures. Here, we demonstrate record-breaking, non-volatile control of optical anisotropy in two-dimensional ReS$_2$ via ferroelectric gating with aluminum scandium nitride (AlScN). Our ferroelectric field-effect transistors achieve near-unity (≈95%) LD tunability of differential reflectance at room temperature—the highest reported for any electrically controlled 2D optical system. Crucially, the programmed optical states exhibit exceptional retention exceeding 12,000 seconds without applied bias, enabling true non-volatile optical memory. Through combined experimental characterization and *ab initio* calculations, we reveal that ferroelectric polarization switching induces substantial asymmetric charge transfer to ReS$_2$, selectively populating conduction band states and triggering structural distortions that dramatically enhance optical anisotropy in the "up" polarization state while leaving the "down" state unperturbed. This ferroelectric-semiconductor coupling provides a universal platform for voltage-programmable, energy-efficient photonic devices with dynamic polarization control, addressing critical needs in integrated photonics as well as programmable far-field optics and telecommunications infrastructure.

**INTRODUCTION**

Optical anisotropy—the directional dependence of light propagation through materials—underpins virtually all modern linear and non-linear photonic technologies, from liquid crystal displays to optical fiber communications and polarized optical imaging by serving as a central operating principle for polarizers, waveplates, and phase-matching elements[1-4]. It arises from birefringence, which is the difference in refractive index parallel and perpendicular to a specific orientation axis. When light passes through a crystal with structural or electronic anisotropy along or across an orientation axis, it induces anisotropic light propagation, providing a means to control the polarization of light. Many conventional bulk materials, including liquid crystals[5], polymers[6,7], metamaterials[8], and oxides[9,10], are well known for their optical anisotropy and have been widely commercialized in polarized photonic devices[11-13]. However, their three-dimensional (3D) bulk nature, rigidity, and limited active polarization control present significant challenges for integration into micro- or nano-scale devices[14].

Two-dimensional (2D) van der Waals materials with intrinsic in-plane anisotropy have emerged as promising alternatives, offering atomically thin active layers with strong optical anisotropy[15-19]. Among them, $ReS_2$ is particularly compelling due to its distorted 1T' crystal structure featuring quasi-one-dimensional Re-Re chains that generate substantial linear dichroism (LD) along specific crystallographic axes[20-22]. Despite extensive research demonstrating LD values up to 30% in pristine $ReS_2$, achieving dynamic, electrically controlled modulation has remained challenging[22]. Previous approaches using conventional gate dielectrics have achieved limited LD tunability (33-46%) in materials like black phosphorus, but critically require large, continuously applied voltages (±80 V) for operation, making them impractical for low-power applications[23]. More fundamentally, these electrostatic approaches provide only volatile control, i.e., the optical state is lost immediately when power is removed, precluding applications requiring optical memory or standby power elimination.

Ferroelectric gate dielectrics offer a transformative solution by providing switchable, remanent polarization that maintains electrostatic fields without continuous power consumption. Wurtzite-phase scandium-doped aluminum nitride (AlScN) represents an ideal ferroelectric gate material, combining exceptionally large remanent polarization ($P_r \approx 80\text{-}115$ μC/cm$^2$), full CMOS compatibility, and demonstrated ability to induce carrier densities exceeding $10^{14}$ cm$^{-2}$ in 2D channels[24-27]. Despite this potential, ferroelectric control of optical anisotropy in 2D materials has not been realized experimentally.

Here, we demonstrate the first non-volatile, electrically programmable control of optical anisotropy using ferroelectric-gated ReS$_2$. Our approach achieves record-high LD tunability (≈95%) at room temperature with retention times exceeding 12,000 seconds, establishing non-volatile control of the fundamental optical property of LD. Through detailed *ab initio* modeling, we reveal the fundamental mechanism: asymmetric charge transfer induced by ferroelectric polarization switching that selectively modulates the electronic structure and atomic configuration of ReS$_2$. Our work opens pathways for voltage-programmable photonic systems with applications spanning optical computing, telecommunications, and polarized imaging.

## RESULTS AND DISCUSSION

**Ferroelectric gating in in-plane anisotropic ReS$_2$.**

**Figure 1a** shows the schematic of a back-gated FeFET device architecture. We use ferroelectric AlScN (100 nm) on Pt/Si substrate as the gate dielectric. ReS$_2$ flakes are exfoliated from a commercially available bulk crystal and transferred onto the AlScN substrate prior to device fabrication. Details of the device fabrication can be found in the Experimental section. **Figure 1b** displays one of the FeFETs studied in this work. **Figure 1c** is a schematic of the ReS$_2$ crystal structure (top view). Single-layer ReS$_2$ consists of a plane Re atoms sandwiched between two S planes. However, unlike other TMDs, strong Re-Re covalent bonds favor conversion from the symmetric trigonal prismatic (2H) structure to antiprismatic (locally distorted ReS$_6$ octahedra, 1T″) coordination, creating pseudo-1D Re-Re chains within the basal plane (along the *b*-axis).[22] (shown by green arrow in **Fig. 1c**). As reported in the literature[28], light can be polarized parallel/perpendicular to this *b*-axis, giving rise to in-plane optical anisotropy.

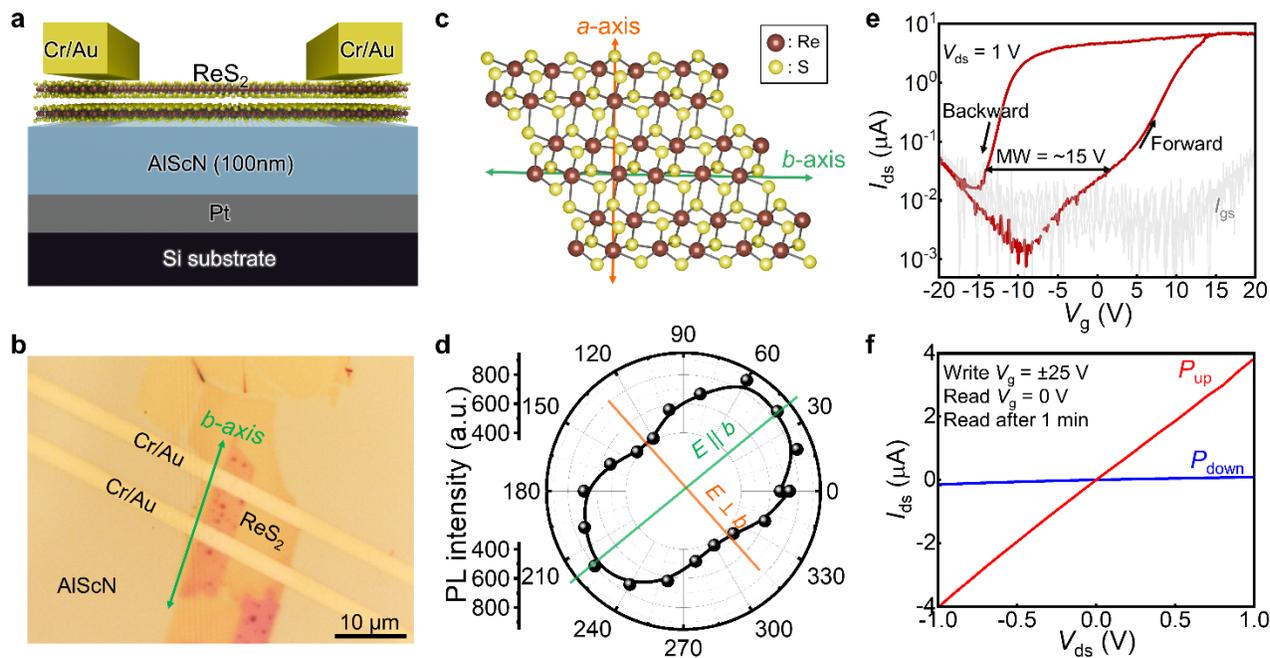

**Figure 1. Non-volatile control of electrical conductivity and optical anisotropy of ReS$_2$.** (**a**) Schematic of the back gate ferroelectric field-effect transistor (FeFET) device architecture showing ferroelectric AlScN on a Pt/Si substrate. ReS$_2$ flakes are exfoliated and then dry transferred onto the AlScN substrate prior to device fabrication. (**b**) Optical image of a representative ReS$_2$/AlScN FeFET. (**c**) Top-view schematic of the ReS$_2$ crystal structure. One-dimensional Re–Re (Red spheres) chains are aligned along the *b*-axis (green arrow), enabling in-plane optical anisotropy (S atoms: yellow). (**d**) Linearly-polarized photoluminescence (PL) spectrum of the ReS$_2$ device shown in (b) measured at room temperature. (**e**) Representative transfer curve of a ReS$_2$/AlScN FeFET illustrating a counterclockwise hysteresis loop with a memory window >15 V as the gate voltage ($V_g$) is swept. (**f**) Output curves for the polarization up ($P_{up}$) and down ($P_{down}$) states, recorded at $V_g$ = 0 V after a ±25 V pulse (applied for 1 s), demonstrating non-volatile and controllable channel conductance modulation.

**Figure 1d** presents linearly polarized PL results of the ReS$_2$ device on AlScN collected at room temperature. Raw PL data for these measurements can be found in **Figure S1**. The PL intensity follows a clear $\cos^2\theta$ dependence on the excitation polarization angle $\theta$, defined as the angle between the incident electric-field vector (*E*) and the crystallographic *b*-axis[22,29,30]. This behaviour reflects the fact that the transition-dipole moment of the lowest-energy excitons ($X_1$, $X_2$) is oriented nearly parallel to the Re–Re chain (*b*-axis); thus, when *E* ∥ *b*-axis ($\theta \approx 40°$ in **Fig. 1d**), the PL reaches its maximum, whereas rotating the polarization by 90° to $\theta \approx 130°$ ($E \perp b$) suppresses the signal[22,30]. The polarization angle that maximizes PL is therefore used as an experimental marker of the *b*-axis, and we set this direction as the reference for all subsequent measurements. As a note, for each device we collect polarized PL separately to determine the $b$-axis of the ReS$_2$ flake prior to conducting gate-controlled LD experiments. Additionally, we use PL measurement primarily to determine the thickness of the flake, as reported in the literature. One such thickness determination of ReS$_2$ flakes can be found in **Fig. S1**. Note that, similar to other 2D TMDs, ReS$_2$ exhibits layer-dependent bandgap (1.4 eV in the bulk to 1.7 eV for the monolayer[31]); however, owing to its weak interlayer coupling, its bandgap remains direct at all thicknesses.

A representative transfer curve of a ReS$_2$/AlScN FeFET is shown in **Fig. 1e**. As the gate voltage ($V_g$) is swept from –20 V to +20 V, the channel drain-source current ($I_{ds}$) changes as the electrical conductivity transitions from the OFF state to the ON state. Furthermore, when $V_g$ is swept in the opposite direction from +20 V to –20 V, the high electrical conductivity is maintained until about –5 V, after which it enters the OFF state again. Therefore, the hysteresis loop of the transfer curve is drawn in a counterclockwise direction, which is typical of FeFET behavior due to ferroelectric gating. The shift in threshold voltage, i.e., memory window (MW), is found to be > 15 V during the DC $V_g$ sweep. For AlScN, when $V_g$ is applied sufficiently negatively (e.g., $V_g$ = –20 V), the ferroelectric polarization inside the AlScN directs downward ($P_{down}$); conversely, when $V_g$ is applied sufficiently positively (e.g., $V_g$ = +20 V), the polarization switches upwards ($P_{up}$), changing the channel conductance. **Figure 1f** shows the output curve of the $P_{up}$ and $P_{down}$ states as $V_g$ (= ±25 V applied for 1 s) is varied (i.e., programmed/erased) and measured at $V_g$ = 0 V after 1 min. As can be seen, the channel conductance is nonvolatile and controllable depending on the $P_{up}$ and $P_{down}$ states.

**Tunable optical anisotropy and its non-volatility**

To demonstrate a programmable ferroelectric gate-tunable LD in ReS$_2$, we perform differential reflectance measurements on ReS$_2$-AlScN devices. **Figure 2a,b** present the linearly polarized optical response of the device shown in **Fig. 1b** under positive bias (switching the AlScN polarization upward, $P_{up}$) and negative bias (switching the polarization downward, $P_{down}$). Here, the green curve shows the response when incoming light is polarized along the ReS$_2$ $b$-axis, and the red curve is for incoming light polarized perpendicular to the ReS$_2$ b-axis. Inset of **Fig. 2a** shows the configuration of initial light polarization and the direction of ReS$_2$ crystal *b*-axis. Details of the deferential reflectance measurements and angle-dependent differential reflectance spectra of this device for $P_{up}$ and $P_{down}$ are displayed in **Fig. S2** and **S3**. LD is calculated using **Equation 1** below[32,33]:

$$(\Delta R_{par} - \Delta R_{per}) / (\Delta R_{par} + \Delta R_{per}) \qquad (1)$$

, where $\Delta R_{par}$ and $\Delta R_{per}$ are the differential intensities of polarized reflectance parallel and perpendicular to the ReS$_2$ *b*-axis, respectively. **Figure 2c,d** display the LD measured for $P_{up}$ and $P_{down}$ state, respectively. As can be seen, for the $P_{up}$ state, we can achieve a very high degree of tunability, with LD above 95% for measurements performed at room temperature at the wavelength of 815 nm (band edge of ReS$_2$). Whereas, for the $P_{down}$ state, LD is ≈20 % at the band edge. Although near-unity LD has been demonstrated in 2D NbOCl$_2$ through its intrinsic in-plane anisotropy in the ultraviolet regime[34] and in 2D antiferromagnetic FePS$_3$ via cavity enhancement in the visible spectrum[35], achieving on-chip LD that is electrically tunable has remained challenging. Our device overcomes this limitation by delivering similarly high LD that can be switched reversibly with a low gate bias.

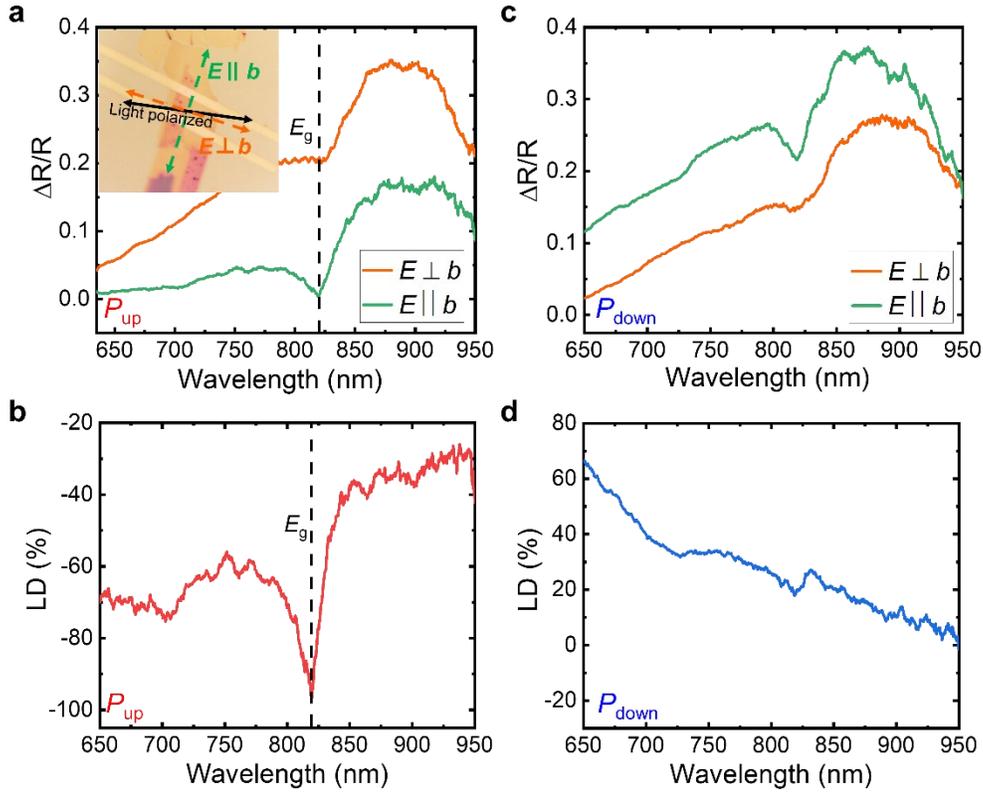

**Figure 2. Linear dichroism (LD) of differential reflectance in ReS$_2$ controlled by AlScN at room temperature.** (**a, b**) Differential reflectance spectra of the ReS$_2$/AlScN device (same as in Fig. 1b) under $P_{up}$ and $P_{down}$ state. In each panel, the green curve represents the signal when the incoming light is polarized along the ReS$_2$ b-axis, while the red curve corresponds to polarization perpendicular to the b-axis. The inset in (a) illustrates the configuration of the incident light polarization and the ReS$_2$ crystal b-axis. (**c, d**) LD under the $P_{up}$ and $P_{down}$ states, respectively. Notably, the $P_{up}$ state exhibits a tunable LD of ≈95% at room temperature at the ReS$_2$ band edge (815 nm), whereas the $P_{down}$ state shows an LD of ≈20%.

To further validate our polarized optical response results, we conduct polarized PL measurements on our ReS$_2$/AlScN devices. **Figure 3a,b** shows the polarization-resolved PL spectrum of the ReS$_2$/AlScN device measured for ferroelectric gate switched ON/OFF states ($P_{up}$-$P_{down}$) measured at room temperature, respectively. The PL peak centered at ≈820 nm (1.51 eV) with a weak shoulder around 780 nm (1.59 eV), a signature of the 5L structure of ReS$_2$ (see **Fig. S1**). To calculate the degree of linear polarization (DoLP) for PL, we use the equation,

$$(I_{par} - I_{per}) / (I_{par} + I_{per}) \qquad (2)$$

, where $I$ stands for PL intensity. As can be seen, PL spectrum collected parallel and perpendicular to the ReS$_2$ crystal $b$-axis in $P_{up}$ state has higher DoLP with ≈44% than $P_{down}$ state with ≈31%, in qualitative agreement to the LD results presented in **Fig. 2**. We also confirm that this trend is reproducible in ReS$_2$ devices of different thicknesses, which show similar DoLP behavior between the $P_{up}$ and $P_{down}$ states (**Fig. S4**).

We also measure polarization-resolved PL spectra of the device for both $P_{up}$ and $P_{down}$ states at the lower temperature of ≈77 K. **Figure 3b,c** displays polarization-resolved normalized PL spectrum for the two ferroelectric switching states, respectively. At 77 K, we can clearly observe two distinct signature PL peaks of ReS$_2$ (labeled as $X_1$ and $X_2$ in **Fig. 3c,d**), both of which exhibit polarization dependency. For the $P_{up}$ state, the DoLP parallel/perpendicular to the ReS$_2$ $b$-axis are determined to be ≈63% for $X_1$ and ≈65% for $X_2$. By contrast, for the $P_{down}$ state, these values are determined to be ≈39% and ≈40%, respectively. A similar trend is also observed in a different device, confirming its reproducibility (**Fig. S6**). As a control experiment, we also measure polarization-resolved PL spectra of ReS$_2$ on a SiO$_2$/Si substrate and compared with the ReS$_2$/AlScN device before and after polarization switching (**Fig. S7**). The DOLP of ReS$_2$ on the AlScN substrate (≈44 % for $X_1$ even before bias) is higher than that observed in the ReS$_2$ crystal on SiO$_2$ (≈32 % for $X_1$), which is attributed to the intrinsic net positive polarization (partial $P_{up}$) of AlScN even before initial poling[24,36] (see **Fig. S7**).

We also observe a stable retention time for the DOLP exceeding 12,000 seconds for both the $P_{up}$ and $P_{down}$ states, measured at 77 K (**Fig. 3e**). These findings hold significant promise for non-volatile optoelectronic memory devices that encode both electrical conductance and emission/extinction polarization. This extended retention time is primarily attributed to the robustness of the ferroelectric polarization state. For example, this long retention is confirmed by the excellent retention of the FeFET channel drain-source current ($I_{ds}$), indicating that once the $P_{up}$ and $P_{down}$ state is written, they remain stable (**Fig. 3f**). Importantly, such long-lived preservation of polarization-resolved optical signals is rarely achieved in anisotropic 2D materials, marking a critical step toward electrically programmable, optically readable non-volatile functionalities in hybrid ferroelectric/2D platforms.

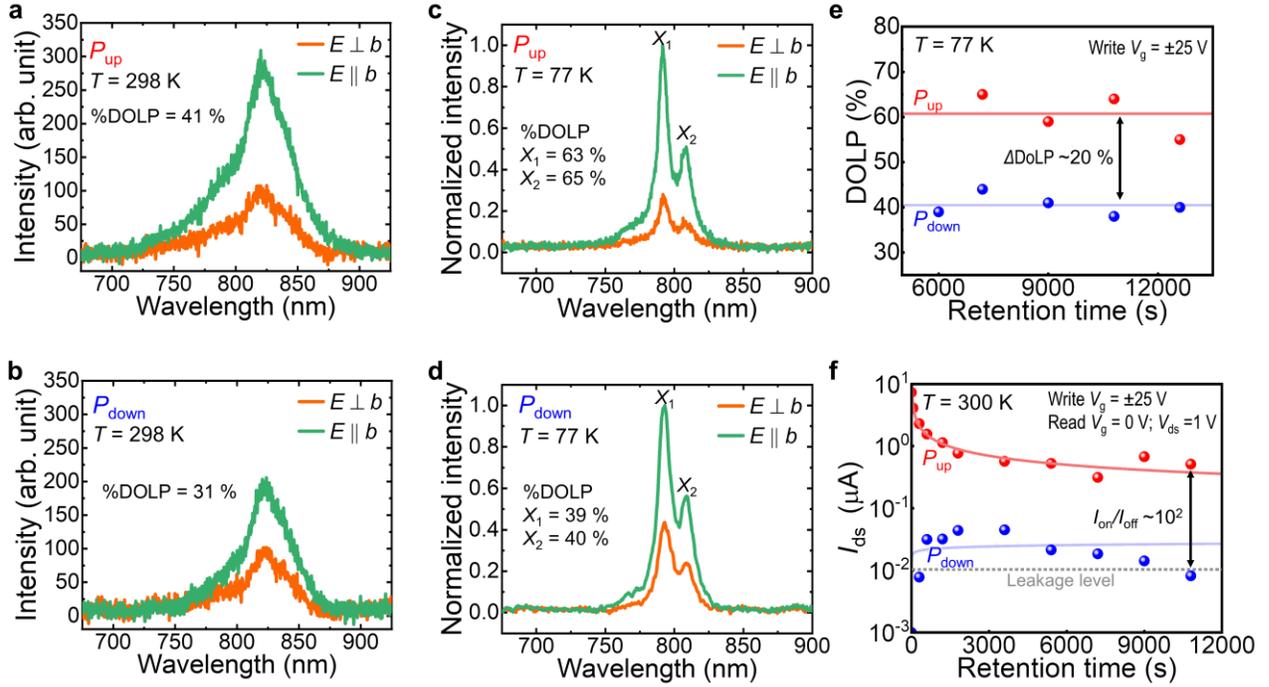

**Figure 3. Polarization-resolved PL spectroscopy and their retention.** (**a, b**) Room-temperature polarization-resolved PL spectra of the ReS$_2$/AlScN FeFET for the (**a**) $P_{up}$ and (**b**) $P_{down}$ states. The degree of linear polarization (DoLP) is ≈44% for the $P_{up}$ state and ≈31% for the $P_{down}$ state. (**c, d**) Normalized polarization-resolved PL spectra measured at ≈77 K for the (**c**) $P_{up}$ and (**d**) $P_{down}$ states. Two distinct PL peaks, labeled $X_1$ and $X_2$, are clearly observed, each exhibiting strong polarization dependence. For the $P_{up}$ state, the DoLP is ≈63% for $X_1$ and ≈65% for $X_2$, whereas for the $P_{down}$ state these values decrease to ≈39% and ≈40%, respectively. (**e**) Polarization-resolved PL measurements reveal a stable retention time for the DOPL measured at 77 K. (**f**) FeFET channel read current ($I_{ds}$) retention, indicating robust stability of the programmed $P_{up}$ and $P_{down}$ states.

**Theoretical analysis of the interface**

A simplified model of the device architecture is used to theoretically model the charge transfer and structural properties of a ReS$_2$/AlN/Pt interface. Factors considered when designing the model system are discussed in **Supplementary Note S1**. The relaxed geometries, the *k*-resolved electronic band structures, and the projected density of states of the interfaces are shown in **Fig. 4**. The band structure and density of states reveal that the $P_{up}$/$P_{down}$ interfaces passivate the polar AlN using different mechanisms. In the $P_{up}$ interface, the ReS$_2$ and AlN have a strong charge transfer interaction, as seen through the closer distance between the layers in the relaxed structure (**Fig. 4c, f**). Examination of the electronic structure shows corresponding behavior, where additional charge accumulates in the ReS$_2$ to passivate the polarization of the AlN substrate. This can be seen in the band structure (**Fig. 4a**), where the lowest conduction band of ReS$_2$ is completely filled by this charge accumulation, and the projected density of states (PDOS) shows that this charge localizes primarily on the Re atoms of the lower ReS$_2$ layer (**Fig. 4b**, and **Fig. S7**), while the layer of ReS$_2$ farther from the interface remains unaffected. The layer projected density of states (**Fig. S8, 9**) suggests that the charge accumulation in the ReS$_2$ layer helps screen the AlN depolarization field in the $P_{up}$ interface, but because the lowest ReS$_2$ conduction band is energetically separated from the higher conduction bands, the passivation is incomplete.

On the other hand, the $P_{down}$ interface shows no major charge transfer to the ReS$_2$ conduction band (**Fig. 4d**). Instead, the electronic structure near the Fermi level is composed primarily of N orbitals (**Fig. 4e**), indicating that holes accumulate on the top AlN surface and that ReS$_2$ is not involved in passivation. In the $P_{down}$ case, the terminating N layer primarily binds holes, and the ReS$_2$ does not significantly stabilize holes to screen the depolarizing field. This shows that the polarization state of the AlN substrate can have a significant impact on the electronic properties of the adjacent ReS$_2$ layer, with the lowest conduction band (CBM) being the most strongly affected.

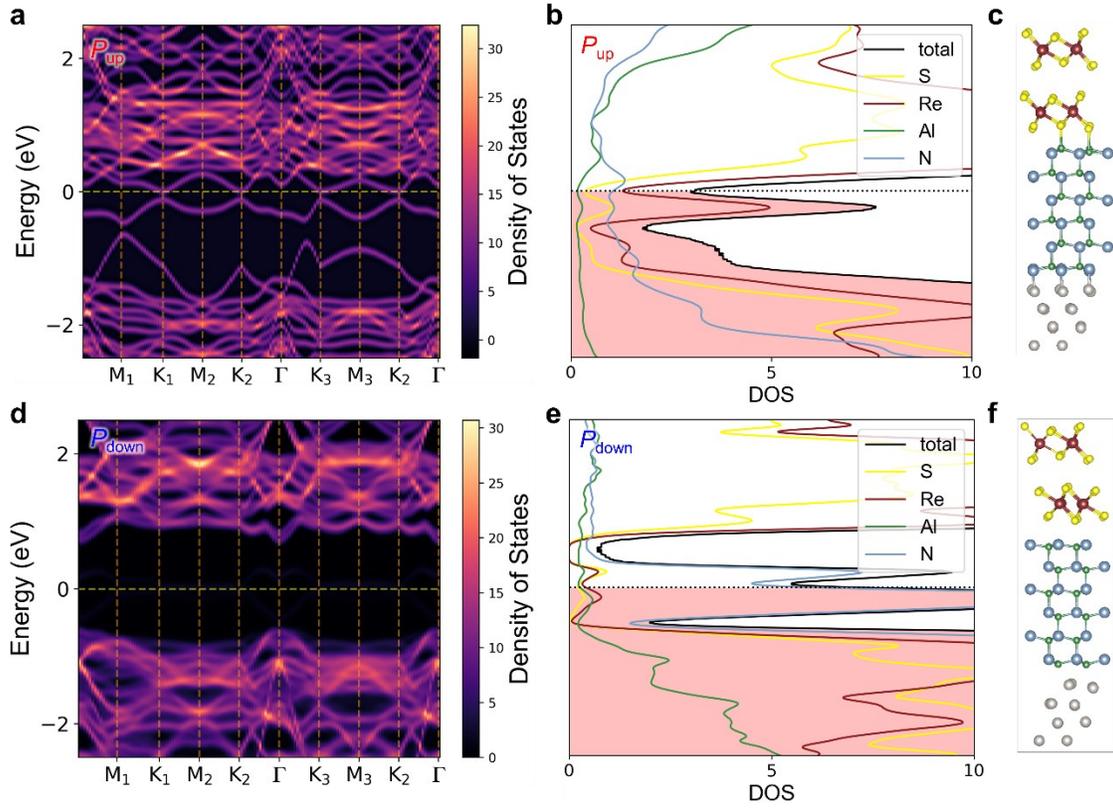

**Figure 4. Electronic properties of the interface structures.** (**a, d**) *k*-resolved density of states projected on ReS$_2$ orbitals for (**a**) $P_{up}$ and (**d**) $P_{down}$ model interfaces. The Fermi level is set to zero and is indicated by a dashed yellow line. (**b, e**) Projected density of states for (**b**) $P_{up}$ and (**e**) $P_{down}$ showing the total orbital contribution of each atomic species. The Fermi level is marked by a black dashed line. (**c, f**) Relaxed geometry of the (**c**) $P_{up}$ and (**d**) $P_{down}$ model interfaces (Re: red, S: yellow, Al: green, N: blue, Pt: gray).

To investigate the effects of charge transfer in the $P_{up}$ interface on the electronic properties of ReS$_2$, the band structure of the ReS$_2$ bilayer is calculated. This is done by calculating the electronic structure (**Fig. 5a**) while preserving the ground state atomic positions of the ReS$_2$ bilayer from the interface (**Fig. 5b**). While the ReS$_2$ structure in the $P_{down}$ interface is barely affected (**Fig. S9**), reflecting its weak interaction with the ferroelectric substrate, the lowest layer of ReS$_2$ in the $P_{up}$ interface undergoes more significant deformation. Here, one out of the two Re–Re pairs in the unit cell separates and the Re atoms move in opposite directions along the *a*-axis. The band structure shows that the CBM is most strongly affected by this distortion (**Fig. 4a**), and the wavefunction corresponding to this band exhibits an anti-bonding character localized on the displaced Re atoms. Peierls distortions have been previously recognized as critical in stabilizing electronic bands in ReS$_2$[31,37], ultimately leading to the formation of the zigzag Re–Re chains along the *b*-axis and the further formation of four-atom diamond shaped clusters. Charge filling of CBM in the $P_{up}$ interface populates anti-bonding states that leads to the ordered dissociation of one pair of Re atoms in adjacent clusters enhancing the Peierls distortion. These insights can be further used to understand the ferroelectric polarization-dependent optical responses for other anisotropic materials.

To connect the structural distortions and charge transfer processes experienced by ReS$_2$ in the interfaces to the optical anisotropy observed in the experiments, we calculate the imaginary dielectric response function of the isolated, distorted ReS$_2$ bilayer (**Fig. 5d-f**). The AlN and Pt layers were deleted, keeping only ReS$_2$ with the charge-induced structural distortions. We make this approximation since the energy range of interest falls within the bandgap of AlN ($\approx$4 eV calculated), and the Pt substrate primarily serves as an electronic reservoir. Moreover, the removal of AlN and Pt resets the charge state of ReS$_2$ with an unfilled CBM. In the interface model, the CBM is fully occupied (**Fig 5a-b**). However, experimental reflectance spectra show that the bandgap of ReS$_2$ in $P_{up}$ vs. $P_{down}$ is relatively unchanged (**Fig. 2**), suggesting that the electronic occupations are not changed enough to allow new optical transitions out of the CBM. This implies that the CBM is only partially filled in the experimental interface, destabilizing the atomic structure but not significantly perturbing the electronic structure. To simulate this, after removal of the AlN and Pt layers, we fix the occupations of the ReS$_2$ so that the conduction band remains empty. Thus, we consider transitions into the CBM and emphasize these transitions as the most important for determining the difference in the optical properties between $P_{up}$ and $P_{down}$.

The dielectric response is computed along two directions: the *b*-axis (parallel to Re chain direction) and the cross direction (perpendicular to the *b*-axis). The results shown in **Fig. 5d-f** reveal anisotropy in the contribution to the dielectric response from the CBM. However, the direction of the anisotropy in the $P_{down}$ and $P_{up}$ structures is different. The $P_{up}$ dielectric function exhibits a large early onset peak (**Fig. 5d**) where the parallel dielectric response is greater than the perpendicular response. In contrast, the $P_{down}$ and pristine bilayers show a larger contribution from the perpendicular response compared to the parallel response. This shows that the structural and electronic changes induced by the charge accumulation in the $ReS_2$ play a central role in the determination of the optical properties.

To sum up, in the $P_{up}$ interface, charge accumulates in the $ReS_2$ in order to passivate the ferroelectric substrate, populating the Re orbitals and perturbing the atomic structure. Conversely, in the $P_{down}$ interface, $ReS_2$ does not play a role in screening the depolarizing field. The population of the conduction band of $ReS_2$ in the $P_{up}$ interface contributes to distortions of the Re atoms along the $b$-axis. The changes to the $ReS_2$ atomic structure introduced by this distortion affect the anisotropy of the optical response. These insights provide a framework for engineering the anisotropic optical properties of $ReS_2$ and similar materials, enabling the fine-tuning of device responses through polarization state control.

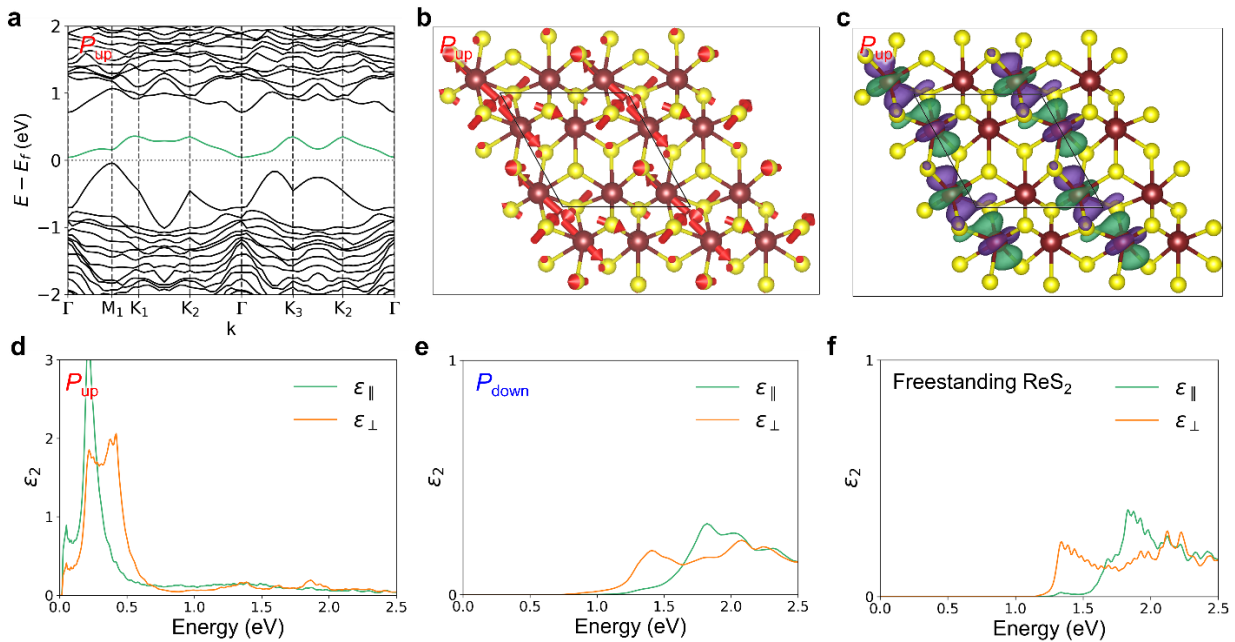

**Figure 5. Theoretical analysis on optical response of $ReS_2$.** (**a-c**) Theoretical analysis on charge induced distortion in $P_{up}$ interface. (**a**) Electronic band structure of the $ReS_2$ bilayer with structural distortions from $P_{up}$ interface highlighting the lowest conduction band (CBM) in green. (**b**) In-plane view (Re: red, S: yellow) of displacement patten of the lowest $ReS_2$ layer in the $P_{up}$ interface relative to the structure of an undistorted bilayer. The outlined black parallelogram shows the unit cell, and the displacements arrows are shown in red. (**c**) Wavefunctions of the lowest conduction band at the Gamma (Γ) point. The positive phase is shown in green, and the negative phase is shown in purple. (**d-f**) The imaginary dielectric function showing only contribution of the lowest conduction band of $ReS_2$ bilayer from (**d**) $P_{up}$ interface, (**e**) $P_{down}$ interface, and (**f**) pristine bilayer. The parallel (∥) direction is along the $b$-axis and the cross direction (⊥) is perpendicular to the $b$-axis.

## CONCLUSION

We demonstrate the first non-volatile, electrically programmable control of optical anisotropy in a 2D material (ReS$_2$) through a back-gated FeFET employing ferroelectric AlScN, achieving record-breaking linear dichroism tunability (≈95%) with retention times exceeding 12,000 seconds. The exfoliated ReS$_2$ exhibits distinct in-plane optical properties due to pseudo-1D Re-Re chains along the *b*-axis, as we confirm through polarization-resolved PL and differential reflectance measurements. The underlying mechanism—asymmetric charge transfer induced by ferroelectric polarization switching—provides fundamental insights into voltage-programmable material properties. Ferroelectric bound charges selectively populate ReS$_2$ electronic states in the $P_{up}$ configuration and trigger structural distortions, while minimal charge transfer to ReS$_2$ ocuurs in the $P_{down}$ state, thereby establishing the correlation between interfacial charge dynamics, structural perturbations, and anisotropic optical responses. Our study highlights the potential of ferroelectric gating for non-volatile, polarization-controlled optoelectronic devices, providing a versatile framework for engineering anisotropic properties in ReS$_2$ and similar 2D semiconductors for advanced, programmable photonic device applications.

## EXPERIMENTAL PROCEDURES

### Device fabrication

$Al_{0.68}Sc_{0.32}N$ with a thickness of 100 nm is deposited on pre-deposited Pt using a pulsed-DC reactive sputtering deposition system (Evatec, CLUSTERLINE 200 II pulsed DC PVD). (See details in Refs.[25-27,38]) The co-sputtering process uses separate 4-inch Al and Sc targets, at a chuck temperature of 350 °C, with 10 sccm of Ar gas flow and 25 sccm of $N_2$ gas flow under the constant pressure of ~1.45 × $10^{-3}$ bar. The preferentially oriented (111) Pt layer promotes the growth of AlScN with highly textured ferroelectricity along the [0001] direction. Few-layer $ReS_2$ thin crystals (flakes) are mechanically exfoliated from the bulk crystal and then transferred onto AlScN/Pt substrates using a PDMS stamp under ambient conditions. Source and drain contact of Cr/Au (10/40 nm) are defined through e-beam lithography (Elionix ELS-7500EX). The Pt beneath the AlScN is used as a gate contact, where the gate contact region was exposed via wet etching of AlScN on the corner of the substrate by dipping it in the KOH solution post fabrication of the Au/Cr/$ReS_2$/AlScN/Pt heterostructure. No additional heat treatment is applied to the sample after device fabrication.

### Device measurements

Room-temperature electrical measurements are performed in air at ambient temperatures in a Lakeshore probe station using a Keithley 4200A semiconductor analyzer. For the devices in **Figs. 2** and **3**, all the optical measurements are conducted after electrical programming/erasing at the probe station. The interval from applying the electrical signal for $P_{up}$ and $P_{down}$ states to conducting the optical measurement (including moving the sample and setting up the apparatus) is ≈5 min. During this transfer of sample, the sample was exposed to room temperature air. Optical measurements are carried out using a Horiba LabRam HR Evolution confocal microscope. A 633 nm (1.96 eV) continuous-wave laser serves as the excitation source for PL, and the white light source (AvaLight-HAL) was used for the reflectance measurement. For polarization-resolved PL and reflectance studies, a half-wave plate is used to control the polarization of linearly polarized excitation light, which is subsequently focused onto the sample. A 50× objective with NA≈0.3 is used both for the sample excitation, and for capturing the emitted signal via an electron-multiplying charge coupled detector. For low-temperature PL measurements at 77 K, the samples are maintained on a cryogenic Linkam stage (THMS600) under a vacuum level of ∼$10^{-3}$ Torr, and the temperature is lowered at a rate of ≈5 °C/min using a liquid nitrogen cryostat.

**Theoretical calculations**

Optimized atomic positions and electronic structures were obtained using density functional theory (DFT) as implemented in the Quantum Espresso software package[39,40]. The exchange correlation energy is treated using the generalized gradient approximation of Perdew-Burke-Ernzerhof (PBE)[41]. A planewave basis set with a cutoff energy of 50 Ry for wavefunctions and 320 Ry for charge density was used, and the core electrons are represented using GBRV ultrasoft pseudopotentials[42]. Grimme's semi-empirical DFT-D3 dispersion correction is used to capture the van der Waals (vdW) interactions[43] between the $ReS_2$ layers. The Brillouin zone was sampled using a $4 \times 4 \times 1$ Monkhorst-Pack mesh. In calculations of the heterogeneous slab, the Fermi surface is treated using gaussian smearing with a value of 0.136 eV, and a dipole correction[44] was used to cancel the electric field in the vacuum region. Geometric relaxation of the interface structures was performed with a force convergence threshold of 0.026 eV/Å and a total energy convergence threshold of 0.0026 eV.

A denser $k$ point grid of $12 \times 12 \times 1$ was used for density of states calculations. Optical spectra were calculated for the $ReS_2$ with structural distortions induced by charge transfer in the interface. The AlN and Pt layers were removed and the $ReS_2$ bilayer distortions were preserved. The occupations were fixed to include transitions into the lowest conduction band. The imaginary dielectric function was computed using momentum matrix elements obtained from the ground-state wavefunctions and charge density of the free standing $ReS_2$ bilayer. Wavefunctions were calculated non-self consistently on a denser $32 \times 32 \times 1$ $k$-point grid using a planewave cutoff of 50 Ry for the valence electrons and optimized norm-conversing pseudopotentials[45,46] generated from OPIUM for the core electrons. A broadening parameter of 0.027 eV was used to obtain high resolution of the dielectric tensor. We calculate $\varepsilon_2$ according to Fermi's golden rule as written in Equation 1.

$$\varepsilon_2(\omega) = \frac{2\pi e^2}{\varepsilon_0 m^2 \omega^2} \int \frac{d\mathbf{k}}{(2\pi)^3} |\hat{\mathbf{a}}_0 \cdot \mathbf{P}_{fi}|^2 \delta[E_{fi}(\mathbf{k}) - \hbar\omega] \quad (1)$$

Where $m$ is the electron mass, $\omega$ is the frequency of incident light, $\hat{\mathbf{a}}_0$ is a unit vector denoting the light polarization, $\mathbf{P}_{fi}$ is an element of the interband momentum matrix (dipole matrix element), and $E_{fi}$ is the energy difference between starting and ending bands $i$ and $f$. A sum over all direct interband transitions between the occupied and unoccupied states was performed.

## SUPPLEMENTAL INFORMATION

Fig S1-9, and Note S1.

## ACKNOWLEDGMENTS

D.J., S.S. and M.R. acknowledge primary support for this work from the Office of Naval Research (ONR) Nanoscale Computing and Devices program (N00014-24-1-2131) and partial support from National Science Foundation (NSF) Future of Semiconductors (FuSe) program award number 2328743. D.J. and K-H.K. acknowledge support from the Air Force Office of Scientific Research (AFOSR) GHz-THz program grant number FA9550-23-1-0391. D.J., J.L. and B.C. acknowledge support from ONR Metamaterials program (N00014-23-1-2037). The materials modeling and theoretical spectroscopy of A.M.S. and A.M.R. were supported by the U.S. Department of Energy, Office of Science, Basic Energy Sciences, under Award No. DE-SC0024942. A.C.K. acknowledges the support of the National Science Foundation through the Graduate Research Fellowship Program (NSF-GRFP). A portion of the sample fabrication, assembly, and characterization were carried out at the Singh Center for Nanotechnology at the University of Pennsylvania, which is supported by the National Science Foundation (NSF) National Nanotechnology Coordinated Infrastructure Program grant NNCI-1542153. Computational support was provided by the National Energy Research Scientific Computing Center (NERSC), a U.S. Department of Energy, Office of Science User Facility located at Lawrence Berkeley National Laboratory, operated under Contract No. DE-AC02-05CH11231. This work was also supported by the Institute for Basic Science, South Korea (IBS-R036-D1), and through the National Research

Foundation (NRF) of Korea (Grant No. RS-2025-00516532) funded by the Ministry of Science and ICT.

**AUTHOR CONTRIBUTIONS**

M.R. and S.S. performed most of the experiments with assistance from B.C., K.-H.K., W.L., J.L., and H.S.S.. A.C.K. and A.M.S. performed the theoretical calculations under supervision of A.M.R. M.R. and S.S. wrote the manuscript, with input from A.C.K. All the authors revised and commented on the manuscript; D.J. and S.S. conceived, planned, and supervised the project. All authors contributed to the writing of the manuscript and interpretation of the data.

**DECLARATION OF INTERESTS**

The authors declare no competing interests.

# Reconfigurable, non-volatile control of optical anisotropy in ReS$_2$ via ferroelectric gating

Mahfujur Rahaman[1,†], Seunguk Song[1,2,3,†,*], Aaliyah C. Khan[4], Bongjun Choi[1], Aaron M. Schankler[4], Kwan-Ho Kim[1], Wonchan Lee[2,3], Jason Lynch[1], Hyeon Suk Shin[3,5], Andrew M. Rappe[4], and Deep Jariwala[1]*

[1]*Department of Electrical and Systems Engineering, University of Pennsylvania, Philadelphia, Pennsylvania 19104, United States*
[2]*Department of Energy Science, Sungkyunkwan University (SKKU), Suwon 16419, Republic of Korea*
[3]*Center for 2D Quantum Heterostructures (2DQH), Institute for Basic Science (IBS), Suwon 16419, Republic of Korea*
[4]*Department of Chemistry, University of Pennsylvania, Philadelphia, Pennsylvania 19104, USA.*
[5]*Department of Chemistry, Sungkyunkwan University (SKKU), Suwon, 16419, Republic of Korea*

[†]These authors contributed equally: Mahfujur Rahaman, Seunguk Song
[*]Correspondence should be addressed. Email to: seunguk@skku.edu (S.S.), dmj@seas.upenn.edu (D.J.)

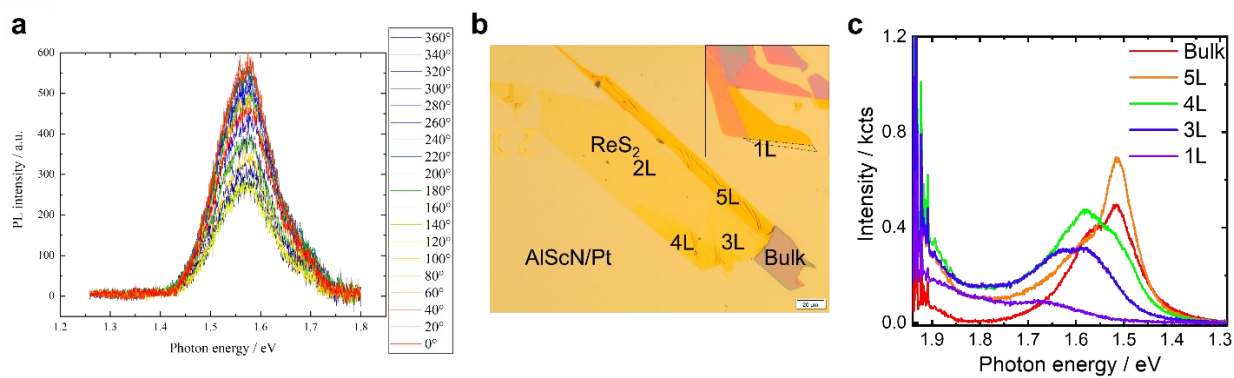

**Figure S1. Raw PL data and flake thickness analysis of mechanically exfoliated ReS$_2$.** (**a**) Raw polarized PL spectra presented in Figure 1d. A 633 nm solid-state laser with 100 µW power is illuminated onto the sample to collect PL signal from the ReS$_2$ device. The light is focused onto the sample using 100x, 0.9 NA objective and the collected light was dispersed using 100 l/mm grove grating. A $\lambda/2$ plate is placed onto the path of excitation to tune the polarization of excited light. Whereas the polarization state of the collected light was kept unchanged, perpendicular to the diffraction grating to have the maximum grating efficiency. The PL peak in centered at 1.57, representative of 3L ReS$_2$. (**b**) Optical image of ReS$_2$ flakes transferred on AlScN. PL measurements are used to determine the thickness of the ReS$_2$ flakes. (**c**) Thickness dependent PL spectra of ReS$_2$ flakes is shown in (b). According to literature[1,2], we can identify the thickness of ReS$_2$ flakes from monolayer to bulk using the PL peak positions.

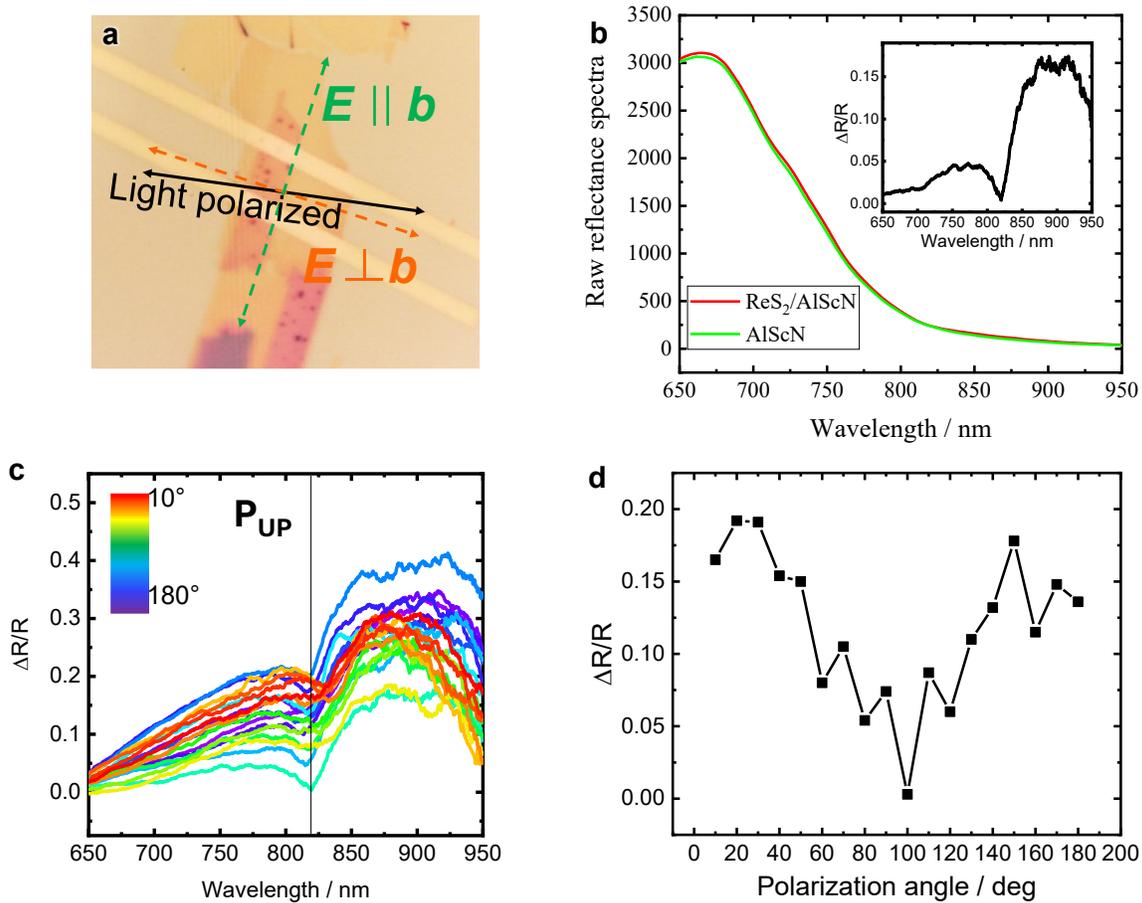

**Figure S2. Angle-dependent differential reflectance spectra under ferroelectric switching.** (**a**) Optical micrograph of the ReS$_2$/AlScN FeFET investigated in Fig. 2a. The crystallographic a-axis (green dashed line) and b-axis (red dashed line) are indicated relative to the incident electric-field vector ($E_0$, black line) at 0° polarizer orientation. The polarizer is then rotated clockwise in 10° steps from 0° to 180°, and a reflectance spectrum was collected at each angle. (**b**) Representative raw reflectance spectra of ReS$_2$/AlScN (orange) and the bare AlScN reference (grey) acquired at a polarizer angle of 100° after applying a positive gate bias. The inset shows the corresponding differential reflectance, $\Delta R/R$. (**c**) Differential reflectance spectra $\Delta R/R$ for all polarization angles (0–180°, 10° step) in the $P_{up}$ state, illustrating the strong anisotropy of ReS$_2$. (**d**) Polar plot of the differential reflectance intensity at 820 nm (near the ReS$_2$ bandgap) as a function of polarization angle, demonstrating tunable optical modulation via polarization control.

**Figure S2a** shows the ReS$_2$ device under study (discussed in **Fig. 2a**). Both ReS$_2$ axis are sketched with respect to the optical field and shown with green and red dashed lines respectively. Incoming optical field (incident light field) at zero degree of the polarizer is shown by the black straight line on the image. We gradually rotate the angle of the polarizer clockwise with a step of 10 degree and collect reflectance spectra for each step. **Figure S2b** presents raw reflectance spectra of ReS$_2$/AlScN and AlScN reference substrate for a polarization angle of 100° measured after the device is positively biased ($P_{up}$ state). To determine the differential reflectance spectra, we use the following formula:

$$\Delta R = \frac{R_{ReS2/AlScN} - R_{AlScN}}{R_{AlScN}}$$

Where $R_{ReS2/AlScN}$ and $R_{AlScN}$ are the polarization angle-dependent raw reflectance spectra. **Inset of Fig. S2b** shows the differential reflectance spectra determined at polarization angle of 100°. Using this method, we determine the differential reflectance spectra of the device for a varying angle between zero degree to 180°. All the Differential reflectance spectra measured under positive bias ($P_{up}$) are displayed in **Fig. S2c**. **Figure S2d** plotted the differential reflectance intensity of the device as a function of polarization angle measured at the ReS$_2$ bandgap at 820 nm. As can be seen, we can modulate the reflectance via tuning the polarization angle of the incident light.

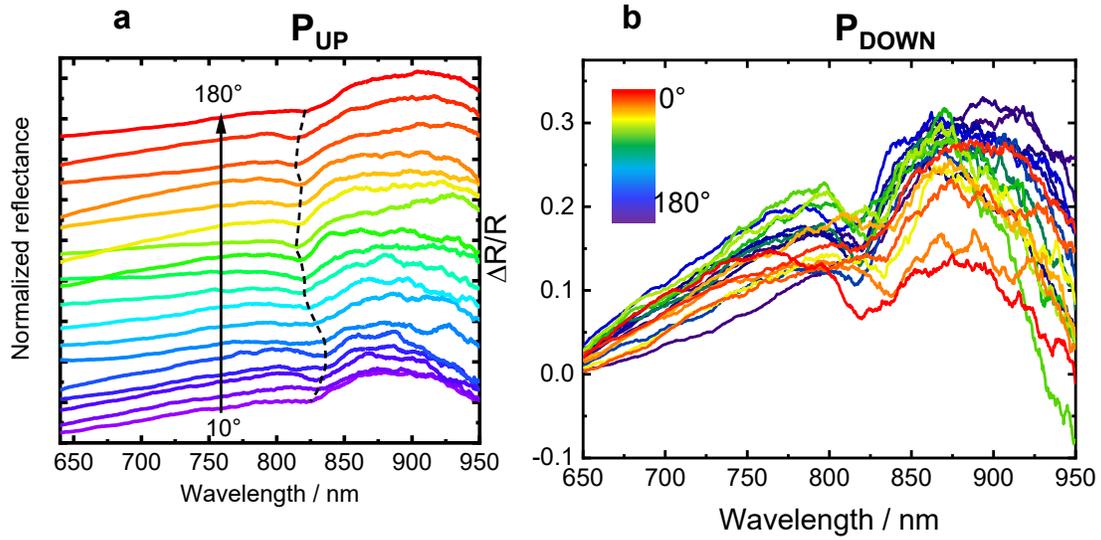

**Fig. S3. Polarization-sensitive exciton response of ReS$_2$ near the band edge.** (**a**) Normalized differential reflectance spectra of the ReS$_2$/AlScN FeFET (from Fig. S2c) under positive bias ($P_{up}$), showing polarization-angle-dependent spectral modulation around the bandgap. Two dominant excitonic features, X$_1$ and X$_2$, exhibit varying spectral weights with polarization. (**b**) Differential reflectance spectra of the same device under negative bias ($P_{down}$), as a function of incident light polarization angle. Similar polarization-sensitive behavior of excitonic features is observed, indicating that the optical anisotropy of ReS$_2$ persists across polarization states.

ReS$_2$ has multiple excitation channels around bandgap. Two prominent excitonic features are X$_1$ and X$_2$ (Ref.[1,2]). As reported in literature, these two excitonic features have different linear polarization sensitivity. As a result, the spectral weight of these excitons varies as the polarization angle of the incident light changes. We also observed similar behavior in our polarization angle dependent reflectance spectra. **Figure S3a** presents normalized differential reflectance spectra of the ReS$_2$ device presented in **Fig. S2c**. As can be seen, with the change of polarization angle of incident light the shape of the reflectance spectra around the bandgap also changes indicative of the change of spectral weight of the excitons X$_1$ and X$_2$. **Figure S3b** displays polarization angle dependent differential reflectance spectra of the ReS$_2$ device discussed in **Fig. S2a** and **Fig. 2** in the main text under negative bias.

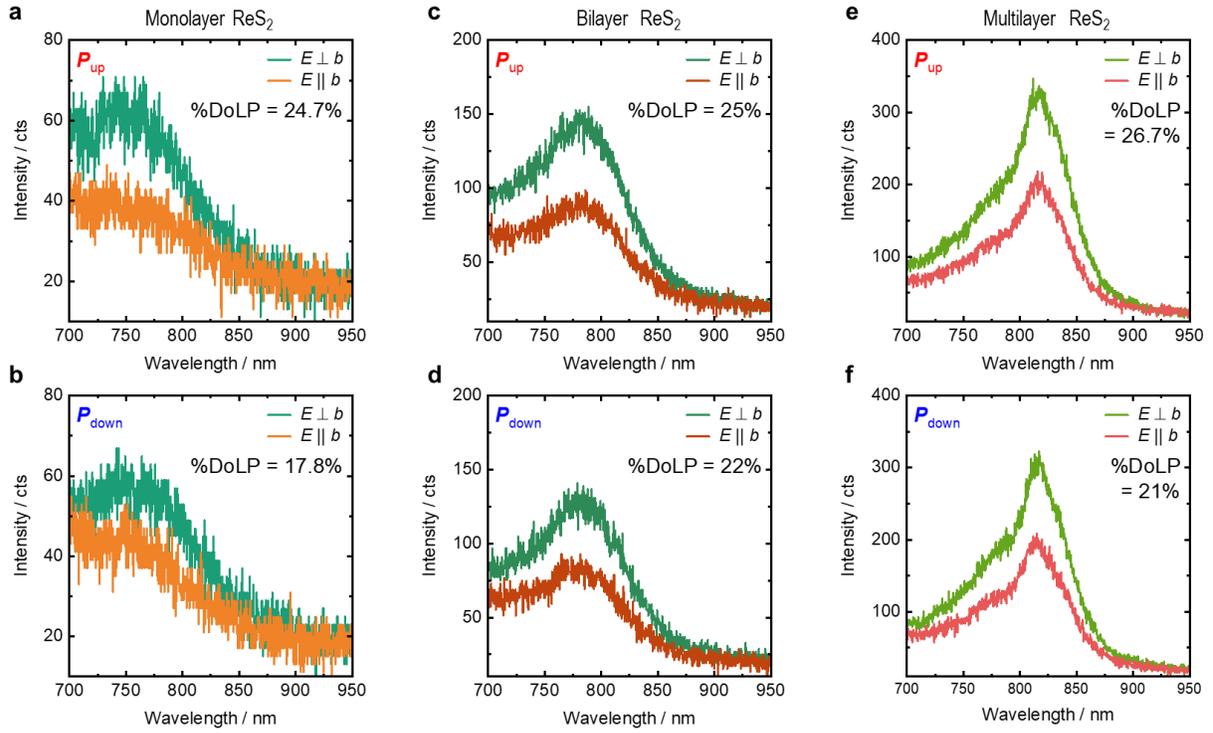

**Figure S4. Thickness-independent DoLP modulation in ReS$_2$ at room temperature.** Monolayer, bilayer, and multilayer devices all exhibit a consistently higher degree of linear polarization in the $P_{up}$ state (~24.7-26.7 %) than in the $P_{down}$ state (~17-22 %), confirming reproducibility across layer numbers.

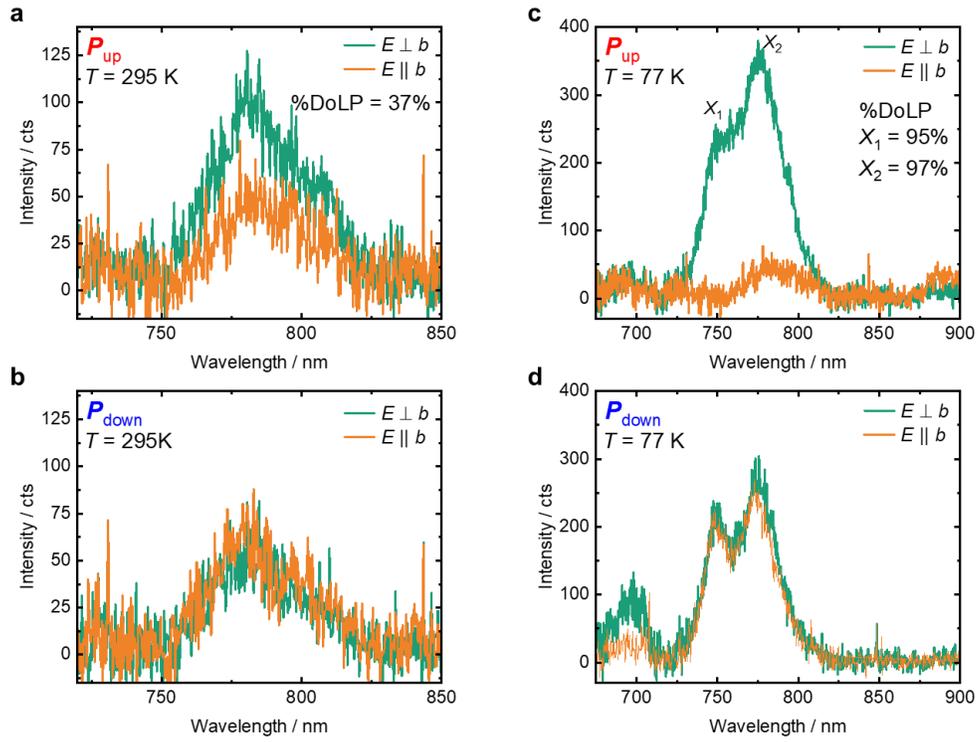

**Figure S5. Polarization-resolved PL spectroscopy.** (**a, b**) Room-temperature polarization-resolved PL spectra of the ReS$_2$/AlScN FeFET for the (**a**) $P_{up}$ and (**b**) $P_{down}$ states. (**c, d**) Normalized polarization-resolved PL spectra measured at ≈77 K for the (**c**) $P_{up}$ and (**d**) $P_{down}$ states.

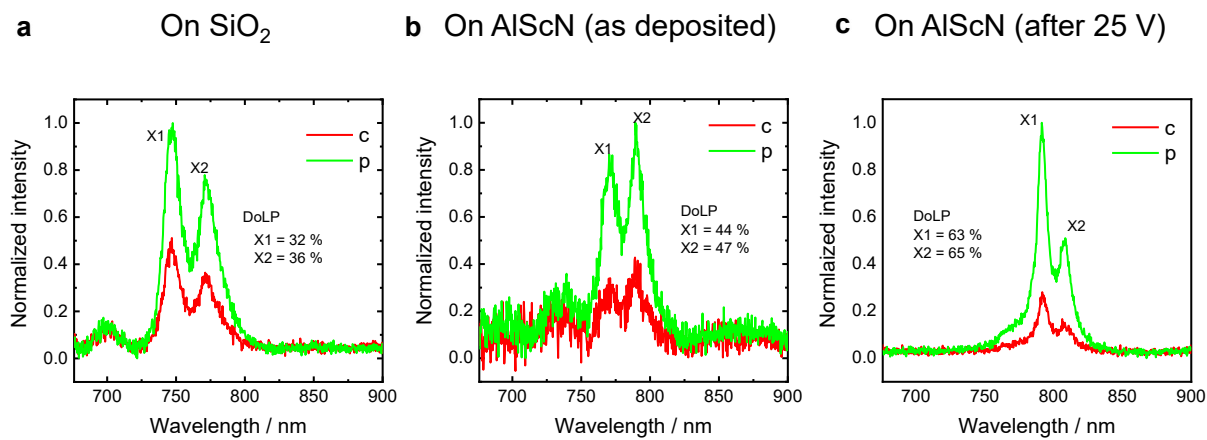

**Figure S6. Substrate-dependent polarization-resolved photoluminescence.** (**a**) Linearly polarized PL spectra of a 5-layer ReS$_2$ flake on a SiO$_2$ substrate at 77 K. Spectra collected with polarization parallel (green) and perpendicular (red) to the ReS$_2$ *b*-axis reveal distinct polarization-dependent excitonic features. (**b**) Linearly polarized PL spectra of an as-deposited 5-layer ReS$_2$ flake on an AlScN under identical measurement conditions, prior to device fabrication. A modest increase in the DoLP is observed compared to the SiO$_2$ case due to the partial polarization of AlScN. (**c**) Linearly polarized PL spectra of the same ReS$_2$ flake after device fabrication under a positive gate bias (corresponding to Fig. 3 in the main text), highlighting further modulation of excitonic polarization response.

**Figure S6a-c** presents comparative study of polarization sensitivity of ReS$_2$ excitons on different substrates. All the PL spectra are collected at 77 K. **Figure S66a** shows the linearly polarized PL spectra of 5L ReS$_2$ flake on SiO$_2$ substrate. Here, green (red) spectra are collected parallel (perpendicular) to the ReS$_2$ *b* axis. **Figure S6b** shows linearly polarized PL spectra of as deposited 5L ReS$_2$ flake on AlScN substrate. The PL spectra presented in **Fig. S6b** are collected before the device fabrication. Comparing the DoLP of ReS$_2$ on SiO$_2$ and AlScN substrates we observed a small increase of DoLP on AlScN substrate for as deposited ReS$_2$ flake. AlScN is well-known for its intrinsically net positive polarization states (i.e., partial polarization)[3]. Therefore, our hypothesis for this small increase of DoLP for as deposited ReS$_2$ on AlScN is that net small positive polarization of AlScN results in small increase in DoLP. For comparison linear polarized PL spectra of ReS$_2$ device discussed in **Fig. 3** in the main text under positive bias are also included in **Fig. S6c**.

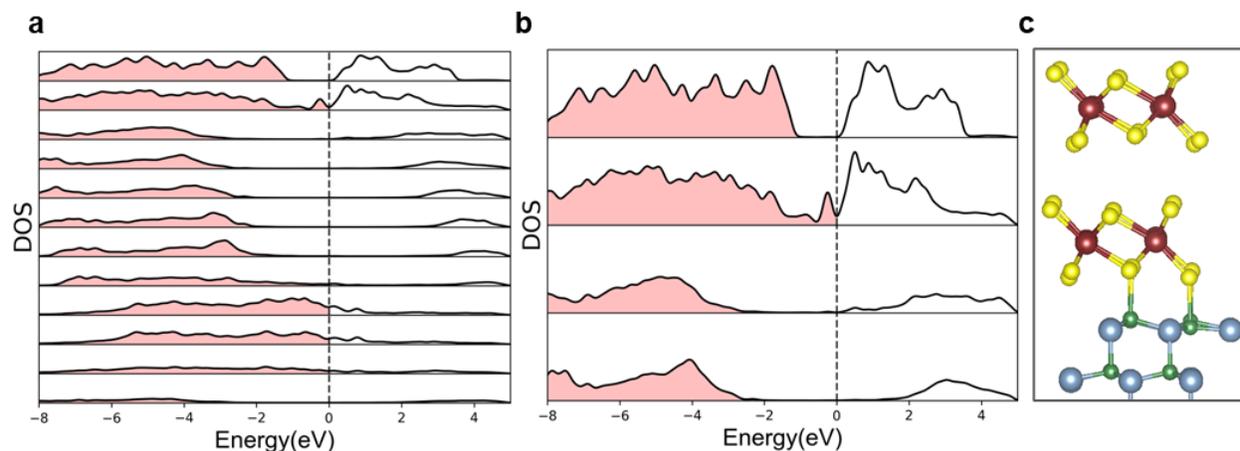

**Figure S7. Layer-resolved electronic structure and interfacial charge transfer in the $P_{up}$ interface.** (**a**) Layer projected density of states (LPDOS) for relaxed $P_{up}$ interface, showing 12 layers going from 4 Pt layers at the bottom, 6 AlN layers in the middle, and two $ReS_2$ layers on top. The Fermi level is set to zero and is indicated by a black dashed line. (**b**) LPDOS of the $ReS_2$ bilayer and top two AlN layers only to show that the top layer of $ReS_2$ is relatively unaffected and charge is transferred only to the layer closest to the AlN. (**c**) Schematic of the $ReS_2$ and interacting AlN layers corresponding to the LPDOS in (b).

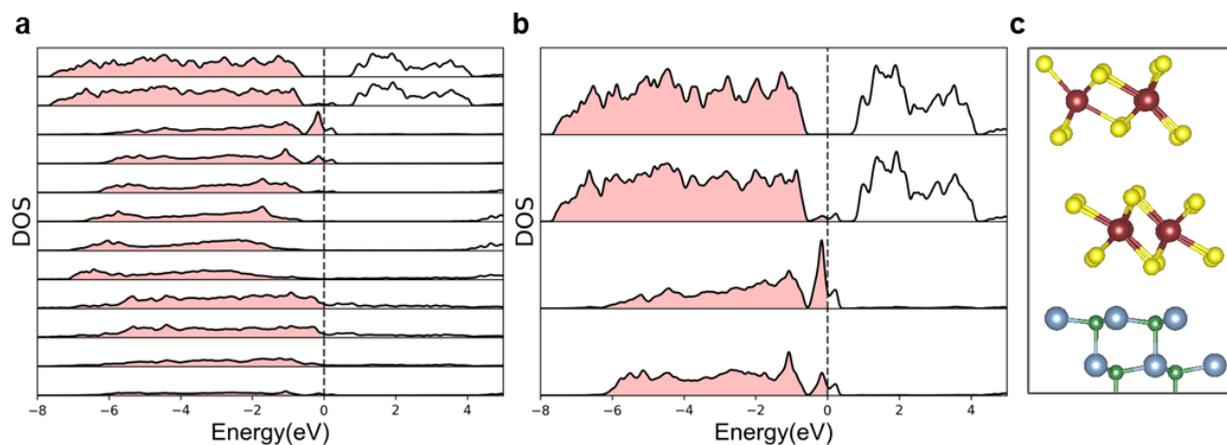

**Figure S8. Layer-resolved electronic structure and interfacial charge transfer in the $P_{down}$ interface.** (**a**) Layer projected density of states (LPDOS) for relaxed $P_{down}$ interface, showing 12 layers going from 4 Pt layers at the bottom, 6 AlN layers in the middle, and two $ReS_2$ layers on top. The Fermi level is set to zero and is indicated by a black dashed line. (**b**) LPDOS of the $ReS_2$ bilayer and top two AlN layers only to show that $ReS_2$ interface layer does not undergo charge transfer. (**c**) Schematic of the $ReS_2$ and AlN layers corresponding to the LPDOS in (b).

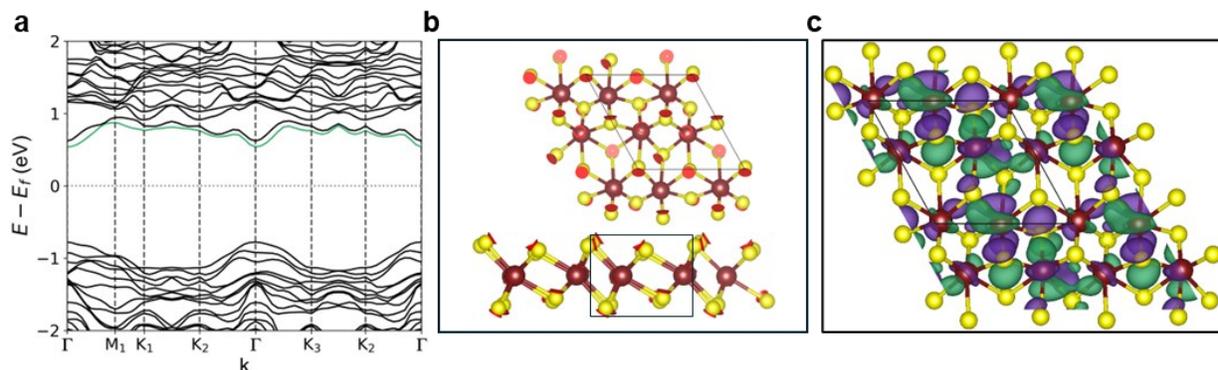

**Figure S9. Theoretical analysis on charge induced distortion in $P_{down}$ interface.** (**a**) Electronic band structure of the ReS$_2$ bilayer with structural distortions from $P_{down}$ interface highlighting the lowest conduction band (CBM) in green. (**b**) In-plane and side views (Re: red, S: yellow) of displacement pattern of the lowest ReS$_2$ layer in the $P_{down}$ interface relative to the structure of an undistorted bilayer. The outlined black parallelogram shows the unit cell, and the displacements arrows are shown in red. (**c**) Wavefunctions of the lowest conduction band at the $\Gamma$ point. The positive phase is shown in green, and the negative phase is shown in purple.

**Supplementary Note S1. Theoretical modeling and interface analysis**

A model system was constructed to theoretically rationalize the observed difference in optical response due to the polarization up ($P_{up}$) and down ($P_{down}$) states of AlScN in the device. A vertically stacked heterostructure was modeled to capture charge transfer and structural changes in the ferroelectric and the $ReS_2$ semiconductor layers of the device. The following guidelines were used in developing a simplified model of the device architecture: (i) Sc doping of AlN serves to lower the coercive field, making ferroelectricity viable in AlN. Therefore, we use pure AlN in our model for simplicity and to accurately represent how the two polar states influence the optoelectronic properties. (ii) The $ReS_2$ layer in the device is much thinner than the ferroelectric layer, so we model a bilayer of $ReS_2$ and 6 layers of AlN. (iii) The modulation of the optical response of $ReS_2$ is primarily due to its interaction with the up and down polarization states of the ferroelectric. We build a simplified model with $ReS_2$ and AlN and add a Pt layer below the AlN to passivate the bottom AlN surface and screen the depolarization field. The final structure was obtained with the following specifications. The topmost layer consists of bilayer $ReS_2$ (001). The middle layer is pure AlN in either its $P_{up}$ or $P_{down}$ polar wurtzite structure. A 6-layer slab of AlN was chosen, as it proved to be sufficient to stabilize the polar state. Finally, a bottom layer of Pt acts as an electron reservoir and screens the depolarization field of the ferroelectric. The in-plane lattice parameters of the AlN, and Pt are matched to those of the $ReS_2$ (001) bilayer which is a rhombohedral cell with $b = 6.40$ Å, $a = 6.5$ Å. The slabs are separated by 20 Å of vacuum to avoid interactions between periodic images. Two total slab models (referred to as interfaces in main text) were created using the $P_{up}$ or $P_{down}$ AlN structures.

**Supplementary References**